\documentstyle[12pt]{article}
\def\hybrid{\topmargin -20pt	\oddsidemargin 0pt
	\headheight 0pt	\headsep 0pt
	\textwidth 6.25in	
	\textheight 9.5in	
	\marginparwidth .875in
	\parskip 5pt plus 1pt	\jot = 1.5ex}

\hybrid

\begin{document}
\def\x{\times}
\def\ra{\rightarrow}
\def\beq{\begin{equation}}
\def\eeq{\end{equation}}
\def\beqa{\begin{eqnarray}}
\def\eeqa{\end{eqnarray}}
\def\D{ {\cal D}}
\def\L{ {\cal L}}
\def\C{ {\cal C}}
\def\N{ {\cal N}}
\def\calE{{\cal E}}
\def\lin{{\rm lin}}
\def\Tr{{\rm Tr}}
\def\mxth{\mathsurround=0pt }
\def\xversim#1#2{\lower2.pt\vbox{\baselineskip0pt \lineskip-.5pt
x  \ialign{$\mxth#1\hfil##\hfil$\crcr#2\crcr\sim\crcr}}}
\def\simgr{\mathrel{\mathpalette\xversim >}}
\def\simle{\mathrel{\mathpalette\xversim <}}

\def\a{\alpha}
\def\b{\beta}
\def\dota{ {\dot{\alpha}} }
\def\lag{Lagrangian}
\def\Kahler{K\"{a}hler}
\def\kahler{K\"{a}hler}
\def\A{ {\cal A}}
\def\C{ {\cal C}}
\def\D{ {\cal D}}
\def\F{{\cal F}}
\def\L{ {\cal L}}
\def\R{ {\cal R}}
\def\x{ \times }
\def\beq{\begin{equation}}
\def\eeq{\end{equation}}
\def\beqa{\begin{eqnarray}}
\def\eeqa{\end{eqnarray}}




\parindent0em

\begin{titlepage}
\begin{center}

\hfill HUB-IEP-94/23\\
\hfill hep-th/9411095\\

\vskip .1in

{\bf ON THE RELATION OF FOUR-DIMENSIONAL
N=2,4 -- SUPERSYMMETRIC STRING BACKGROUNDS
TO INTEGRABLE MODELS}

\vskip .2in

{\bf Gabriel Lopes Cardoso and
Dieter L\"{u}st }  \\
\vskip .1in

{\em  Humboldt Universit\"at zu Berlin\\
Institut f\"ur Physik\\
D-10115 Berlin, Germany}\footnote{e-mail addresses:
GCARDOSO@QFT2.PHYSIK.HU-BERLIN.DE, LUEST@QFT1.PHYSIK.HU-BERLIN.DE}


\end{center}

\vskip .2in

\begin{center} {\bf ABSTRACT } \end{center}
\begin{quotation}\noindent

In this letter we discuss the relation of four-dimensional, $N=2$
supersymmetric string backgrounds to integrable models. In particular
we show that  non-K\"ahlerian gravitational backgrounds with one
$U(1)$ isometry plus
non-trivial antisymmetric tensor  and dilaton  fields arise as the
solutions of the Liouville equation or, for the case of vanishing
central charge deficit, as the solutions of the continual Toda
equation. When performing an Abelian duality transformation, a particular
class of solutions of the continual Toda equation leads to the well-known
gravitational  Eguchi-Hanson instanton background with self-dual
curvature tensor.

\end{quotation}
\vskip 2.0cm
November 1994\\
\end{titlepage}
\vfill
\eject

\newpage

Recently, the investigation of  four-dimensional non-trivial
backgrounds for consistent string propagation has
attracted a lot of
attention \cite{aben,KiKoLu,nw2}.
This research is essentially the continuation
of the long-existing program of finding solutions to vacuum Einstein
gravity in general relativity in the sense that in string theory
a specific form of matter, namely
the dilaton ($\Phi$) plus antisymmetric tensor field ($B_{\mu\nu}$) system,
is added. Of particular interest, in general relativity as well as in string
theory, are those background configurations which exhibit some Killing
symmetries. The reasoning for this is twofold. First, the existence
of symmetries often implies that the background field equations are exactly
solvable since the symmetries effectively reduce the
theory to a lower dimensional
system. In this way the
field equations are very often related to integrable
systems as demonstrated for the case of pure gravity in many examples.
Second, the isometries can be used to generate new solutions starting
from a given solved model. Particular famous examples in general
relativity are the Ehlers and Geroch transformations \cite{ehl}
which relate different
gravitational backgrounds to each other.
In string theory, these transformations are
extended in the sense that they generically mix the gravitational background
with the $\Phi$, $B_{\mu\nu}$ matter system. Specifically, for the case of
an Abelian $U(1)^d$ isometry, the string solution generating transformations
are elements of the group $O(d,d,R)$ \cite{kiri,Verl,giro}. In fact,
one can even formulate a string Geroch group
as a current algebra extension
\cite{bak1} containing as its `zero mode' part not only the
$O(d,d)$ transformations but also  the $S$-field
duality transformations \cite{filq}.
Moreover one can show that
in string theory  the discrete subset of $O(d,d,Z)$ transformations,
the socalled target space duality transformations
\cite{dual}, leave the underlying
conformal field theory invariant. Therefore the duality transformations
generically relate geometrically or
even topologically \cite{giki} different
background spaces which are nevertheless equivalent from the
string point of view.

Of particular importance are those backgrounds which are consistent with
(extended) world sheet supersymmetries and allow
for the construction of an (extended) superconformal algebra. Thus
they provide consistent backgrounds for heterotic or type II superstrings
and are expected to lead also to (extended) supersymmetries in the target
space-time. In \cite{KiKoLu} a systematic discussion on
four-dimensional backgrounds
with $N=2$ world sheet supersymmetry was given.
There, a set of conditions, imposed by world-sheet $N=2$ supersymmetry,
was derived
 for finding K\"ahlerian as well as non-K\"{a}hlerian four-dimensional
(non-compact) target spaces.
For example, a broad class of non-K\"ahlerian target spaces with torsion
could be constructed as solutions to a very simple integrable system,
namely the one
having the Laplace equation as field equation.
In addition, these backgrounds have vanishing central charge
deficit $\delta c$, and one
therefore expects to be dealing with an enhanced $N=4$ superconformal
symmetry. Specifically
the socalled axionic instantons \cite{rey}, which
satisfy a self-duality equation in
the dilaton plus axion matter system, fall into this class. The
dual of the axionic instantons are given by a pure gravitational,
K\"ahlerian background having the form of a cosmological
string \cite{cos}. Also
the gravitational wave plus $B_{\mu\nu}$ background of ref.\cite{nw2}
can be shown \cite{kkl}
to follow from the Laplace equation, and the corresponding currents
of the $N=4$ superconformal algebra can be explicitly constructed \cite{kkl}.

In the following,
a new $N=2$ supersymmetric four-dimensional solution of the non-K\"ahlerian
type, i.e. with non-trivial $B_{\mu\nu}$ and $\Phi$, is presented.
Due to its $U(1)$ isometry this new solution is again related
to an integrable model, namely it
is shown to satisfy an equation of the Liouville type and, in the case of
a vanishing central charge deficit, $\delta c=0$,
also
the continual Toda equation
recently discussed in \cite{Ba} in the context of  self-dual,
purely gravitational backgrounds.
For a vanishing central charge deficit, $\delta c=0$,
this new solution describes a
four-dimensional
target space which is dual to a Ricci flat gravitational background
 describing an
Eguchi-Hanson instanton.
Since it is known that the Eguchi-Hanson background
is consistent with $N=4$ world sheet supersymmetry, this then
provides strong evidence that also the original non-K\"ahlerian background
with $\delta c=0$ is $N=4$ supersymmetric (assuming that the
$N=4$ world sheet supersymmetry
doesn't get destroyed by
$T$-duality transformations).

We begin with a brief review of some of the relevant aspects
presented in \cite{KiKoLu} for the construction of non-trivial
four-dimensional backgrounds with torsion.
The most general $N=2$ superspace action
for one chiral superfield $U$ and
one twisted chiral superfield $V$
in two dimensions is determined by a single real function
$K(U,\bar U,V,\bar V)$ \cite{Leg}
\beqa
S={1\over 2\pi \alpha '}\int{\rm d}^2xD_+D_-\bar D_+\bar D_-
K(U,\bar U,V,\bar V)
\label{action}
\eeqa
The fields $U$ and $V$ obey chiral and twisted chiral
constraints given, respectively, by
\beqa
\bar D_\pm U=0,\qquad\bar D_+V=D_-V=0
\label{constraint1}
\eeqa

The target space interpretation of the theory can be made manifest by
writing down the purely bosonic part of the superspace action
(\ref{action})
\beqa
S =-{1\over 2\pi\alpha '
}\int{\rm d}^2x \lbrack K_{u\bar u}\partial^a u
\partial_a\bar u-K_{v\bar v}\partial^a v\partial_a\bar v
+\epsilon_{ab}(K_{u\bar v}\partial_a u\partial_b\bar v
+K_{v\bar u}\partial_a v\partial_b\bar u)\rbrack
\label{bosonic}
\eeqa
where
\beqa
K_{u\bar u}={\partial^2K\over\partial U\partial\bar
U}\;\;,\;
K_{v\bar v}={\partial^2K\over\partial V\partial\bar V}\; \;,\;
K_{u\bar v}={\partial^2K\over\partial U\partial\bar V}\;\;,\;
K_{v\bar u}={\partial^2K\over\partial V\partial\bar
U}
\label{metrics}
\eeqa
Here,
$u$ and $v$ denote the lowest components of the superfields $U$
and $V$.
Thus, the first two terms in (\ref{bosonic})
describe, in complex coordinates, the metric background of the model
given by
\beqa
G_{\mu\nu}=\pmatrix{0&K_{u\bar u}&0&0\cr K_{u\bar u}&0&0&0
\cr 0&0&0&-K_{v\bar v}\cr 0&0&-K_{v\bar v}&0\cr}.
\label{metric}
\eeqa
To obtain a space with Euclidean signature, one has to require
$K_{u\bar u}$ to be positive definite whereas
$K_{v\bar v}$ has to be negative definite.
Note that the metric (\ref{metric}) is not K\"{a}hler.

The $\epsilon_{ab}$-term in (\ref{bosonic}) provides the
antisymmetric
tensor field background
\beqa
B_{\mu\nu}=\pmatrix{0&0&0&K_{u\bar v}\cr
0&0&K_{v\bar u}&0
\cr 0&-K_{v\bar u}&0&0\cr -K_{u\bar v}&0&0&0\cr}.
\label{bmn}
\eeqa
It follows that the field strength $H_{\mu\nu\lambda}$
\beqa
H_{\mu\nu\lambda}
=\partial_\mu B_{\nu\lambda}
+\partial_\nu B_{\lambda\mu}
+\partial_\lambda B_{\mu\nu},\label{hfield}
\eeqa
can also be expressed entirely in terms
of the function $K$ as
\beqa
H_{u\bar u v}
=
{\partial^3K\over\partial U\partial\bar U\partial V}\;,\;
H_{u\bar u\bar v}
=-
{\partial^3K\over\partial U\partial\bar U\partial \bar V} \;,\;
H_{v\bar v u}
=
{\partial^3K\over\partial V\partial\bar V\partial U}\;,\;
H_{v \bar v \bar u}
=-
{\partial^3K\over\partial V \partial\bar V \partial \bar
U}
\label{torsion}
\eeqa

In order to have
that the non-K\"{a}hlerian background given by (\ref{metric})
and ({\ref{bmn})
provides a
consistent string solution, the string equations
of motion have to be satisfied.  These are obtained by requiring the
vanishing of the $\beta$-function equations, which at the one-loop level are
given by \cite{ft}
\beqa
0&=&\beta_{\mu\nu}^G=R_{\mu\nu}-{1\over
4}H_\mu^{\lambda\sigma}
H_{\nu\lambda\sigma}+2\nabla_\mu\nabla_\nu\Phi +O(\alpha') \nonumber\\
0&=&\beta_{\mu\nu}^B=\nabla_\lambda
H_{\mu\nu}^\lambda-2(\nabla_\lambda
\Phi )H_{\mu\nu}^\lambda+O(\alpha')
\label{betaf}
\eeqa
$\Phi(u,\bar u, v, \bar v)$ denotes the dilaton field.
The central charge deficit $\delta c$ for the gravitational
background  is determined by the vanishing of the
$\beta$-function of the dilaton
field as
\beqa
\delta c\equiv c-{3D\over 2}={3\over 2}\alpha'
\lbrack 4(\nabla\Phi)^2-4\nabla^2\Phi-R+{1\over 12}
H^2\rbrack+O(\alpha'^{2})
\label{dilaton}
\eeqa
$\delta c$ must be zero in order that the $N=2$ world sheet supersymmetry is
extended to $N=4$. Then, in case of $N=4$,
the one-loop $\beta$-functions do not get higher
order corrections and the lowest order solution is an exact superconformal
field theory.

The equations of motion (\ref{betaf}) and (\ref{dilaton})
for the non-K\"{a}hlerian background
(\ref{metric}) and (\ref{bmn})
were explicitly
worked out in \cite{KiKoLu}.  There, it was found
that the following
quantities, ${\cal U}= ln K_{u {\bar u}}$ and ${\cal V}= ln K_{v {\bar v}}$,
have to satisfy the following
set of differential equations
\beqa
\partial_u {\cal V}&=& 2\partial_u\Phi+\bar C_1(\bar u)e^
{\cal U}  \nonumber\\
\partial_v {\cal U} &=&2\partial_v\Phi+\bar C_2(\bar v)e^{\cal V}
\label{b}
\eeqa
Here, $\bar C_1(u)$ and $\bar C_2(v)$ denote two apriori arbitrary
holomorphic functions which arise as integration "constants" when solving
the $\beta$-function equations (\ref{betaf}).  In order to proceed with the
solving of (\ref{betaf}) and (\ref{dilaton}), the following exclusive
cases were then considered:
(i) $C_1=C_2=0$; (ii) $C_1=0$, $C_2\neq 0$; (iii) $C_1,C_2\neq 0$.
We will, in the following, only be concerned with
case (ii).  It is for this case that we will construct a new non-trivial
solution, which for a vanishing central deficit (\ref{dilaton}) will turn
out to be the dual of the Eguchi-Hanson instanton.

As shown in \cite{KiKoLu}, when considering case (ii) it is
useful to perform the following holomorphic
change of coordinates
\beqa
w=\int{{\rm d}v\over C_2(v)}
\label{changew}
\eeqa
Then, it can be shown that (\ref{betaf}) implies that
${\cal V}=
{\cal V}(u,\bar
u,w+\bar w)$,
${\cal U}=
{\cal U}(u,\bar u,w+\bar w)$ as well as
$\Phi=\Phi(u,\bar u,
w+\bar w)$.
Thus, case (ii) necessarily leads to at
least one
$U(1)$ Killing symmetry.  The string equations of motion (\ref{betaf})
 can then for case (ii) finally be shown
to be equivalent to the following differential equation for the K\"{a}hler
potential $K(u,\bar u,w + \bar w)$
\beqa
 \partial_w e^{K_w}
=e^{c_1(w+\bar w)+c_2} K_{u\bar u}
\label{solveb}
\eeqa
as well as to
\beqa
2\Phi= ln K_{w \bar w} - c_1(w+\bar w) + constant
\label{solvea}
\eeqa
Equation (\ref{solvea})
defines the dilaton field $\Phi=\Phi(u,\bar u, w + \bar w)$.
$c_1$ and $c_2$ denote two arbitrary constants.
Finally, the central charge deficit (\ref{dilaton}) is proportional to the
constant
$c_{1}$ appearing in (\ref{solveb}), namely
\beqa
\delta c = - 2 c_1
\label{deltac}
\eeqa

We now proceed to construct a non-trivial solution of (\ref{solveb})
for the K\"{a}hler potential $K$.  Once this is achieved, it is then
straightforward to obtain the associated four-dimensional gravitational
background through (\ref{metric}), (\ref{bmn}) and (\ref{solvea}).
We begin by making an ansatz for $K_{u \bar u}$ as follows
\beqa
K_{u \bar u} = \Upsilon (u, {\bar u})\; \Omega (w+{\bar w})
\label{kuu}
\eeqa
As we will see, this separation ansatz reduces the field equations
to a two-dimensional differential equation for $\Upsilon$.
Then, integrating (\ref{kuu}) with respect to $u$ and ${\bar u}$ yields
\beqa
K= \Omega(w + {\bar w}) \int du d{\bar u} \Upsilon(u, {\bar u})
+ h(w+ {\bar w})\, ( z(u) + {\bar z}({\bar u})) + r(w+ {\bar w})
\label{intuu}
\eeqa
where the functions
$h(w+ {\bar w}),  z(u)$ and $r(w+ {\bar w})$ denote integration
"constants".
Inserting (\ref{kuu}) into (\ref{solveb}) and integrating once with
respect to $w + {\bar w}$, on the other hand, yields
\beqa
K= a(u, {\bar u}) + \int\limits^{w+{\bar w}}
ln \left(f(u,{\bar u}) + \Upsilon(u,{\bar u})
\int\limits^{w+{\bar w}}
\, e^{c_1(w+{\bar w})+c_2} \, \Omega(w+{\bar w})\right)
\label{intw}
\eeqa
where the functions
$a(u, {\bar u})$ and
$f(u,{\bar u})$
denote two additional integration "constants".
Equating (\ref{intuu}) and (\ref{intw}) shows that compatibility of both
can be achieved by demanding that
\beqa
f(u,{\bar u}) = \alpha \Upsilon (u, {\bar u})
\label{fupsi}
\eeqa
where $\alpha$ denotes an arbitrary constant.
Then, (\ref{intw}) turns into
\beqa
K= a(u, {\bar u}) + (w+{\bar w})\; ln \Upsilon (u,{\bar u})
+ \int\limits^{w+{\bar w}}
ln \left( \alpha +
\int\limits^{w+{\bar w}}
\, e^{c_1(w+{\bar w})+c_2} \, \Omega(w+{\bar w})\right)
\label{modw}
\eeqa
Hence,
it follows by inspection that
\beqa
r(w+{\bar w}) = \int\limits^{w+{\bar w}}  ln \left(\alpha
+ \int\limits^{w+{\bar w}}
\, e^{c_1(w+{\bar w})+c_2} \, \Omega(w+{\bar w})\right) &&\nonumber\\
a(u,{\bar u}) +
(w+{\bar w})\; ln \Upsilon (u,{\bar u}) = \Omega(w+{\bar w})
\int du d{\bar u} \Upsilon (u, {\bar u})
+ h(w+\bar{w})\; (z(u)+{\bar z}({\bar u})) &&
\label{rw}
\eeqa
There are now two possibilities.  Either $\Omega=const$ or $\Omega \neq
const$.  We will in the following take $\Omega \neq const$.  Then, it follows
from (\ref{rw}) that
\beqa
a(u,\bar u)&=&0 \nonumber\\
(w+{\bar w})\; ln \Upsilon (u,{\bar u}) &=& \Omega(w+{\bar w})
\int du d{\bar u} \Upsilon (u, {\bar u})
+ h(w+\bar{w})\; (z(u)+{\bar z}({\bar u}))
\label{auu}
\eeqa
Now, differentiating (\ref{auu}) with respect to $u$ and $\bar u$ yields
\beqa
(w+{\bar w}) \; \partial_u \partial_{\bar u} ln \Upsilon(u, {\bar u}) =
\Omega(w+{\bar w}) \Upsilon(u,{\bar u})
\label{liouv}
\eeqa
Thus, it follows from (\ref{auu}) and (\ref{liouv}) that
\beqa
\Omega(w+{\bar w}) &=& \gamma \; (w+\bar w) \nonumber\\
h(w+{\bar w}) &=& \tau \; (w+\bar w)
\label{ome}
\eeqa
where $\gamma$ and $\tau$ are two arbitrary constants.
It then also follows from (\ref{liouv})
that $ln \Upsilon$ has to satisfy the Liouville equation
\beqa
\partial_u \partial_{\bar u} ln \Upsilon(u, {\bar u}) =
\gamma \, \Upsilon(u,{\bar u})
\label{liouville}
\eeqa
The general solution to the Liouville equation (\ref{liouville}) is well
known  and reads
\beqa
\Upsilon (u, {\bar u}) = \frac{\partial_u F(u)
 \partial_{\bar u} G({\bar u})}{(1-\frac{\gamma}{2} FG)^2}
\label{solliou}
\eeqa
where $F(u)$ and $G({\bar u})$ are arbitrary functions of $u$ and ${\bar u}$,
respectively.
Acting with $SL(2,C)$ transformations
on the functions $F(u)$ and $G({\bar u})$
leaves $\Upsilon$ invariant and therefore
does not change the background.

Thus, we have constructed a non-trivial solution to (\ref{solveb})
for the K\"{a}hler potential $K$.  $K$ can
now be directly
read off from (\ref{modw}) and reads
\beqa
K= (w+{\bar w})\; ln \Upsilon (u,{\bar u})
+ \int\limits^{w+{\bar w}}
ln \left( \alpha  + \gamma
\int\limits^{w+{\bar w}}
\, e^{c_1(w+{\bar w})+c_2} \, (w+{\bar w})\right)
\label{kap}
\eeqa
where, again, $\Upsilon$ is the solution (\ref{solliou})
to the Liouville equation.
The dilaton field takes the following form
\beqa
2\Phi=\log\biggl(\frac{\gamma(w+\bar w)}{\alpha  + \gamma
\int\limits^{w+{\bar w}}
\, e^{c_1(w+{\bar w})+c_2} \, (w+{\bar w})}
\biggr) + constant
\label{dilw}
\eeqa

In the following, we will now specialise to the case where the central
charge deficit (\ref{deltac}) vanishes, that is when $c_1=0$.  Then the
theory
is expected to possess $N=4$ world sheet supersymmetry. We will
also set $c_2= i \pi$.
  Then, the K\"{a}hler
potential (\ref{kap}) is evaluated to be
\beqa
K=
 (w+{\bar w})\; ln \Upsilon (u,{\bar u})
+ \int\limits^{w+{\bar w}}
ln \left( \alpha - \frac{\gamma}{2}
 (w+{\bar w})^2 \right)
\label{kapc1}
\eeqa
Due to the Liouville nature of $\Upsilon$ it can now
be easily shown that the
differential equation (\ref{solveb}) for the K\"{a}hler potential
(\ref{kapc1})
is
related to the continual Toda equation recently discussed in \cite{Ba},
as follows.\footnote{Actually, ref.\cite{Ba}
 discussed the continual Toda equation for
 K\"ahlerian backgrounds with self-dual metric.}
  Introducing $\Sigma=K_w$, it follows from (\ref{kapc1}) and
 (\ref{liouville})
that
\beqa
\partial_u \partial_{\bar u} \Sigma = \gamma \,\Upsilon(u,{\bar u})
\eeqa
Differentiating (\ref{solveb}) once with respect to $w + {\bar w}$
then indeed yields the continual Toda equation
\beqa
\partial_w^2 e^{\Sigma} = -
\partial_u \partial_{\bar u} \Sigma
\label{toda}
\eeqa

The line element associated with (\ref{kapc1})
is readily computed to be
\beqa
ds^2 &=&  - K_{w{\bar w}} \, dw d{\bar w} + K_{u {\bar u}} \, du d{\bar u}
\nonumber\\
&=& \frac{ \gamma (w+{\bar w})}{ \alpha - \frac{
\gamma}{2} (w+ {\bar w})^2}
\, dw d{\bar w}
+\gamma  (w+{\bar w})\, \frac{\partial_u F(u)
 \partial_{\bar u} G({\bar u})}{(1-\frac{\gamma}{2} FG)^2}
\, du d{\bar u}.
\label{fgmet}\eeqa
The corresponding scalar curvature only depends on the coordinates $w,\bar w$
since in the $u,\bar u$ `direction' the curvature is constant due to the
Liouville nature of our solution. However the space is asymptotically flat
only for $\alpha=0$. The
antisymmetric tensor field strength
has for example the following
non-vanishing component
\beqa
H_{u\bar uw}=K_{u\bar u w}=\gamma {F'G'\over (1-{\gamma \over 2}FG)^2}.
\label{hexpl}
\eeqa
This shows that, whenever the choice of $F$ and $G$ is such that
$G={\bar F}$, one can use $F$ and $\bar F$ as coordinates of the
four-dimensional background instead of $u$ and $\bar u$. We will now
indeed set $G={\bar F}$.  Then, one can set $F=\sqrt{2} u$ without loss
of generality.  We will, for later convenience,
also set $\gamma=-1$ and
$\alpha = -\frac{\rho^2}{2}$.  Then,
the resulting K\"{a}hler potential reads
\beqa
K&=&  (w+{\bar w}) \; ln2 - 2
 (w+{\bar w})\; ln (1 + u{\bar u})
+ \int\limits^{w+{\bar w}}
ln \left(- \frac{\rho^2}{2} + \frac{1}{2}
 (w+{\bar w})^2 \right) \nonumber\\
&=&  (w+{\bar w}) \left( - 2
-2 ln (1 + u{\bar u})
+
ln (-\rho^2 +
 (w+{\bar w})^2) \right)
+ 2 \rho \,artanh(\frac{w+\bar w}{\rho})
\label{kapu1}
\eeqa
Finally, performing the change $w \rightarrow -w$ yields
\beqa
K=  (w+{\bar w}) \left(  2
+2 ln (1 + u{\bar u})
-
ln (-\rho^2 +
 (w+{\bar w})^2) \right)
- 2 \rho \,artanh(\frac{w+\bar w}{\rho})
\label{kapu}
\eeqa

We now proceed to show that the dual geometry, obtained from the
K\"{a}hler potential dual to
(\ref{kapu}),  describes an Eguchi-Hanson instanton.
\footnote{This dual geometry
will also exhibit the SL(2,C) symmetry of the original geometry}
It is important to emphasize that the duality transformation will not destroy
the $N=2$ world-sheet supersymmetry of the original model, since
it can be performed within the $N=2$ superfield formalism.
The dual K\"{a}hler potential,  $\tilde{K}(u,{\bar u},\psi+{\bar \psi})$,
is obtained by performing an abelian duality transformation \cite{Bu}
with
respect to the U(1) Killing symmetry of the K\"{a}hler potential
$K(u, {\bar u}, w + {\bar w})$.  Alternatively,
the dual K\"{a}hler potential $\tilde{K}(U,{\bar U},\Psi + {\bar \Psi})$
can also
be obtained \cite{Verl} by a Legendre
transformation of $K(U,{\bar U}, W + {\bar  W})$,
which amounts to replacing the twisted chiral
superfield $W$ (whose lowest component field is $w$) by a chiral superfield
$\Psi$ (whose lowest component field is $\psi$), as follows
\cite{Leg}.
 Starting from
\beqa
\tilde K(w+\bar w,u,\bar u,\psi+\bar\psi)=
K(u, \bar u,w+\bar w)-(w+ \bar w)(\psi+\bar\psi)
\label{dualpot}
\eeqa
and demanding that
\beqa
\partial_w \tilde{K} = \partial_w K - (\psi + {\bar \psi}) =0
\label{kw}
\eeqa
allows one to solve for $w+{\bar w}
= g(u,\bar u, \psi,{\bar \psi})$ as a function of $u, \bar u, \psi$
and $\bar \psi$.  Then, reinserting $w+{\bar w}
= g(u,\bar u, \psi,{\bar \psi})$ into (\ref{dualpot}) yields the
dual K\"{a}hler potential $\tilde{K}(u,\bar u,\psi,\bar \psi)$.
The dual background metric is then given by
\beqa
\tilde{G}_{\mu\nu}=\pmatrix{0&\tilde{K}_{u\bar u}&0&\tilde{K}_{u\bar \psi}
\cr \tilde{K}_{u\bar u}&0&\tilde{K}_{\psi\bar u}&0
\cr 0&\tilde{K}_{\psi \bar u}&0&\tilde{K}_{\psi\bar \psi}\cr
\tilde{K}_{u\bar \psi}&0&\tilde{K}_{\psi\bar \psi}&0\cr}.
\label{dualmetric}
\eeqa
The dual antisymmetric tensor field background vanishes, since the dual
K\"{a}hler potential $\tilde{K}$ is now only given in terms of  chiral
superfields.  Finally, the dual dilaton field is given as
\beqa
2 \tilde{\Phi} = 2 \Phi - ln 2 K_{w\bar w}
\label{dualdil}
\eeqa
For the K\"{a}hler potential (\ref{kapu})
under consideration,
it follows from (\ref{solvea}) that the dual dilaton is constant,
$\tilde{\Phi}=const$.  Thus, the dual geometry associated with the dual of
(\ref{kapu})
is entirely described by a purely gravitational background with
Ricci flat background metric (\ref{dualmetric}), which we will compute in
the following.
Inserting the K\"{a}hler potential (\ref{kapu})
into (\ref{kw}) gives $w+\bar w=g(u,\bar u,\psi, \bar \psi)$ as
\beqa
w + \bar w = \sqrt{
 \rho^2 + e^{-(\psi + \bar \psi)} \, (1 + u \bar u)^2 }
\label{guu}
\eeqa
Then, inserting (\ref{guu}) into (\ref{dualpot}) yields the
dual K\"{a}hler potential as
\beqa
\tilde{K}(u,\bar u, \psi,\bar \psi) =
2 g(u,\bar u,\psi, \bar \psi) - 2 \rho\, artanh\frac{g(u,\bar u ,\psi,
\bar \psi)}{\rho}
\label{kadual}
\eeqa
Inspection of (\ref{guu}) shows that a
natural set of holomorphic coordinates is given by
\beqa
z_1 &=& e^{\frac{-\psi}{2}}\nonumber\\
z_2 &=& e^{\frac{-\psi}{2} } \, u
\label{holoco}
\eeqa
such that
\beqa
g(z_1,z_2)= \sqrt{\rho^2 +(z_1 {\bar z}_1 +
z_2 {\bar z}_2)^2}
\label{gzz}\eeqa
After setting
\beqa
Q=z_1 {\bar z}_1 +
z_2 {\bar z}_2
\eeqa
one computes that
\beqa
\frac{d\tilde{K}}{dQ}= 2 \frac{\sqrt{\rho^2 + Q^2}}{Q}
\eeqa
which indeed describes the K\"{a}hler potential \cite{Bon} for the
usual hyperk\"{a}hler Eguchi-Hanson metric.
The background metric  associated with (\ref{kadual}) is,
in coordinates (\ref{holoco}), then
given by \cite{Bon}
\beqa
g_{1 \bar 1}= 2 \left( \frac{g}{Q^2} \,|z_2|^2 + \frac{1}{g} \,|z_1|^2 \right)
\;,\;\;
g_{2 \bar 2}= 2 \left( \frac{g}{Q^2} \,|z_1|^2 + \frac{1}{g} \,|z_2|^2 \right)
\;,\;\;
g_{1 \bar 2}= - 2 \frac{\rho^2}{g\,Q^2} \,z_2 {\bar z}_1
\eeqa
Since this selfdual
 metric is hyperk\"ahler, the $N=4$ world-sheet supersymmetry
is guaranteed.

In summary, we have exhibited the
relation of four-dimensional $N=2,4$ supersymmetric,  non-K\"ahlerian
superstring backgrounds with one Abelian Killing symmetry
 to a particularly simple integrable system, namely the
Liouville equation respectively continual Toda equation for the case
of vanishing central charge deficit. The solutions
of the Liouville resp. Toda
equations
lead to a new class of
metric, $B_{\mu\nu}$ and $\Phi$ backgrounds which
can by explictely constructed. Via the duality transformation,
a particular subset of these solutions is equivalent to the well-known
Eguchi-Hanson instanton in four dimensions. We believe that it is
interesting to further investigate the relation of
integrable systems to non-trivial (super) string backgrounds.

\vspace*{.3in}

{\bf Acknowledgement}

We would like to thank
Ioannis Bakas and Elias Kiritsis and also Christian Preitschopf
for fruitful discussions.

\end{document}